\documentclass[useAMS,letters]{mn2e}
\usepackage{graphicx}
\usepackage{epstopdf}
\def\gsim{\mathrel{\raise.5ex\hbox{$>$}\mkern-14mu
             \lower0.6ex\hbox{$\sim$}}}
\def\lsim{\mathrel{\raise.3ex\hbox{$<$}\mkern-14mu
             \lower0.6ex\hbox{$\sim$}}}

\newcommand{\aap}{    {\it Astron. Astrophys.}}

\newcommand{\apj}{    {\it Astrophys. J.}}

\newcommand{\mnras}{  {\it Mon. Not. Roy. Astron. Soc.}}

\title[EHT]{The immediate environment of an astrophysical black hole}
\author[Contopoulos]{I. Contopoulos$^{1,2}$\thanks{e-mail: icontop@academyofathens.gr}\\
$^{1}$Research Centre for Astronomy and Applied Mathematics,
Academy of Athens, Athens 11527, Greece\\
$^{2}$National Research Nuclear University, 31 Kashirskoe highway, Moscow 115409, Russia}

\begin{document}

\date{Accepted ... Received ...; in original form ...}

\pagerange{\pageref{firstpage}--\pageref{lastpage}} \pubyear{2017}

\maketitle

\label{firstpage}

\begin{abstract}
In view of the upcoming observations with the Event Horizon Telescope (EHT), we present our thoughts on the immediate environment of an astrophysical black hole. We are concerned that two approximations used in general relativistic magnetohydrodynamic numerical simulations, namely numerical density floors implemented near the base of the black hole jet, and a magnetic field that comes from large distances, may mislead our interpretation of the observations. We predict that three physical processes will manifest themselves in EHT observations, namely dynamic pair formation just above the horizon, electromagnetic energy dissipation along the boundary of the black hole jet, and a region of weak magnetic field separating the black hole jet from the disk wind.
\end{abstract}


\section{Expectations from the EHT}

The Event Horizon Telescope (hereafter EHT) is an international collaboration whose primary goal is to image the immediate environment of the two supermassive black holes with the largest apparent event horizons, the $4\times 10^6 M_\odot$ one at the center of our Galaxy in Sagittarius A, and the $7\times 10^9 M_\odot$ one in M87 (see eventhorizontelescope.org for details and information). EHT opens the possibility of imaging extreme light lensing near the horizon, as well as the dynamic evolution of black hole accretion, by pushing the limits of global millimeter and submillimeter VLBI. The first observations of Sagittarius A were obtained in April 2017, but the publication of the first images will require several more months of reduction and analysis.

In this work, we would like to express our concern about the interpretation of the EHT observations with respect to the morphology of accretion and jet formation near the black hole horizon. The EHT Team employs the most comprehensive current theoretical and numerical models of general relativistic (GR) magnetohydrodynamic (MHD) accretion and ejection in order to account for the expected morphology, spectra, and variability of the source (e.g. Lu et al. 2016; Chan et al. 2015 and references therein). These fall into two general categories: models in which a large scale magnetic field of a definite polarity is advected with the accretion flow and generates large scale winds and jets, and models with a random magnetic field component introduced just to trigger the magneto-rotational instability that generates accretion but no large scale ejection flows.

The following three physical processes have not yet been included in these numerical models and need to be taken into account in the interpretation of the observations:
\begin{enumerate}
\item Dynamic electron-positron pair formation at the base of the black hole jet right above the black hole horizon,
\item Electromagnetic energy dissipation along the last open flux surface through the black hole horizon, and 
\item Reversal of the large scale magnetic field polarity outside the black hole jet.
\end{enumerate}

In the following sections, we will discuss how the above three physical processes might affect the observations of the two supermassive black holes at the centers of our Galaxy and M87.

\section{Dynamic pair formation}

GR MHD simulations of black hole accretion consistently yield a region above the black hole horizon, where magnetic forces are balanced by the gravity of the black hole (e.g. Takahashi et al. 1990; Pu et al. 2016). As a result, the flow is separated into inflow and outflow, and this region is quickly (at light crossing times) emptied of material. When that takes place, i.e. when the local density drops below a certain threshold that the code cannot handle, MHD codes implement so-called density floors. The employment of density floors in this region (so-called stagnation or separation surface) is {\em tantamount to a continuous artificial supply of material for the jet above the black hole horizon}. The simulations do not yield anything special in the region that is emptied of material, and a low density funnel of the same material as the rest of the material of the accretion flow, namely electron-proton/ion plasma, forms along the magnetic field lines that cross the black hole horizon. 

In reality, however, as the region above the black hole horizon is emptied of the accretion flow material, it is left only with the accumulated magnetic field and the electric fields associated with black hole rotation. This configuration is akin to that in the pulsar magnetosphere. It is common belief in the study of pulsar magnetospheres (e.g. Goldreich \& Julian~1969) that whenever the magnetosphere is emptied of plasma, strong electric fields develop along the magnetic field, accelerating electrons and positrons to relativistic energies sufficient to generate electron-positron pairs through various astrophysical processes. These pairs cascade down to more and more pairs that quickly (at light crossing times) fill the magnetosphere with a force-free electron-positron ({\em not} electron-proton/ion) plasma. The very same process has been proposed in the original paper on the extraction of matter and energy from a rotating black hole \cite{BZ77} to take place also in black hole magnetospheres. Several researchers have identified the region of pair formation (which is indeed the source of the black hole jet) to be associated with the zero space-charge surface at roughly $2r_{\rm g}$, where $r_{\rm g}\equiv GM/c^2$, and $M$ is the black hole mass (e.g. Ptitsyna \& Neronov~2016). Notice that the expected region of pair formation has nothing to do with the stagnation surface of GR MHD simulations.

Particle formation and acceleration along magnetic field lines that thread the black hole horizon have been considered by members of the EHT Team (e.g. Mo\'{s}cibrodzka et al.~2011). Other teams have also started to discuss this issue (e.g. Pu et al.~2017). Moreover, the study of the dynamics of electrostatic gaps in force-free magnetospheres is very limited, even in the case of pulsars 
(e.g. Timokhin~2010).
Overall, we expect that the source of the jet in the magnetospheric gaps will be sporadic and intermittent, with a variability timescale on the order of the light crossing time $2r_{\rm g}/c$. This is equal to about 1~minute in Sagittarius A and about 1~day in M87, thus only the latter is discernible with a typical VLBI imaging experiment that lasts a couple of hours.


We conclude this section with the prediction that the funnel region above the horizon will be analogous to a pulsar magnetosphere,
with pair formation and injection at its base right above the horizon. This is very different from what GR MHD simulations surmise, namely non-thermal electron-proton/ion injection higher up along the stagnation surface resulting in both an outflow to large distances and an inflow toward the black hole (see e.g. fig.~2 in Pu et al.~2017). In analogy to the origin of the pulsar wind in the pulsar magnetosphere (as manifested in the rich sub-pulse structure and variability), the electron-positron (and not electron-proton/ion) black hole jet will not be continuous and will never be completely filled. It will instead be sporadic and intermittent (see cartoon in Fig.~1). In the case of M87, this variability may be observable with the EHT. 

\begin{figure}
\centering
\includegraphics[width=1.0\columnwidth]{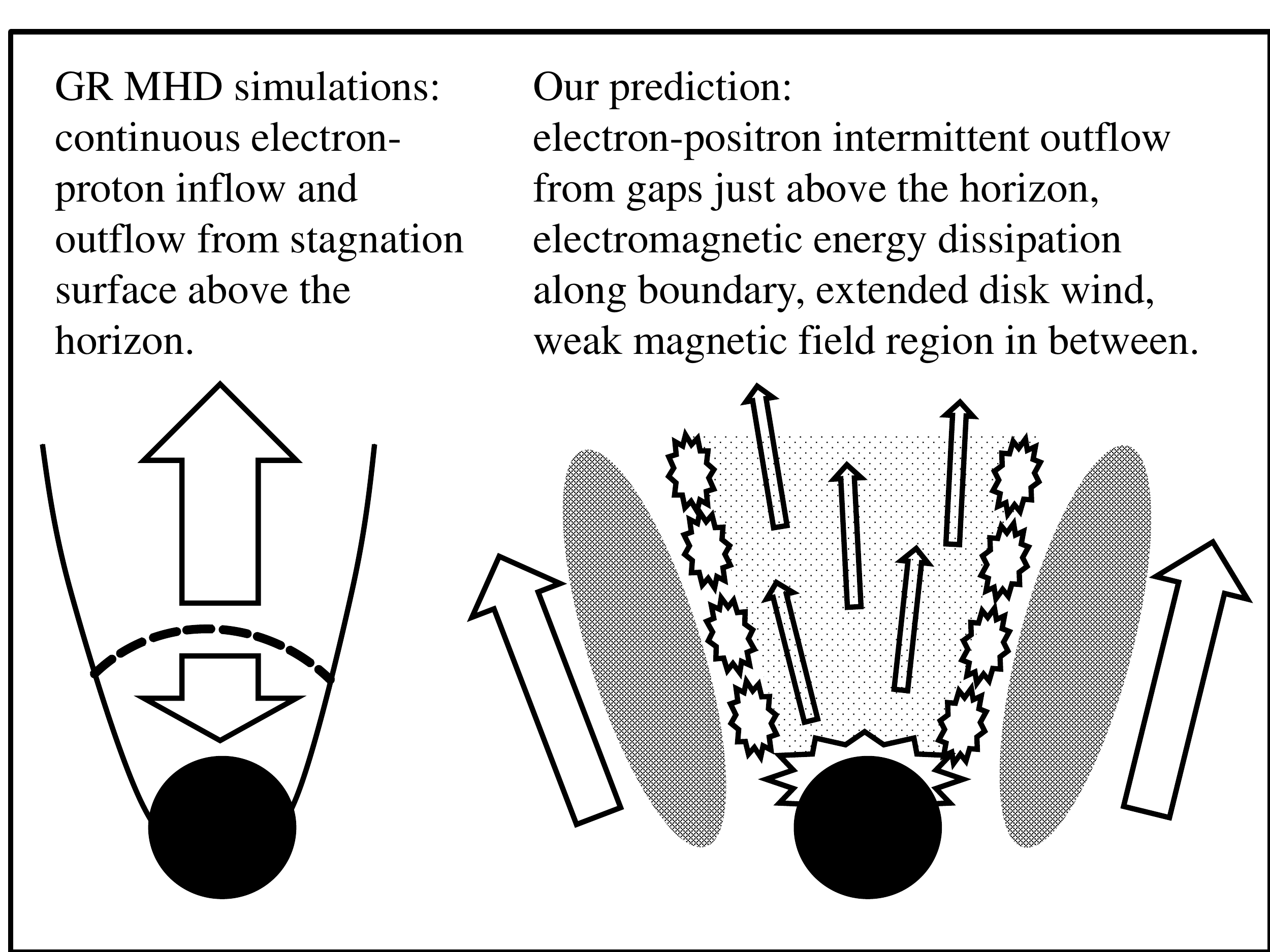}
\caption{Sketches of our prediction (right) and the prediction of GR MHD simulations (left). Black circle: black hole. Dashed line: stagnation surface. Dotted region: electron-positron black hole jet. Everywhere else: electron-proton plasma. Stars: particle acceleration, high energy emission. Grey region between the black hole jet and the extended disk wind: weak magnetic field, low radio emission.}
\label{figure1}
\end{figure}

\section{Electromagnetic energy dissipation}

The second physical process that we would like to emphasize is the development of a strong current sheet along the last open magnetic flux surface that crosses the black hole horizon. This is a fundamental feature of black hole and pulsar magnetospheres. Nathanail \& Contopoulos~(2014) showed that the structure of the steady-state axisymmetric black hole magnetosphere is determined by the condition of smooth crossing of the two light cylinders that appear in the problem, namely the outer one analogous to the pulsar light cylinder, and the inner general relativistic one inside the black hole ergosphere. These are the two Alfven surfaces of the flow, one for the negatively/positively charged outflow outside the zero space charge surface, and one for the positively/negatively charged inflow inside for black hole rotation that is aligned/counter-aligned with the magnetic field respectively. The presence of two light cylinders imposes two restrictions which determine uniquely two unknown distributions of the problem, that of the electric current along open magnetic field lines, and that of the angular velocity of rotation of the magnetic field. In other words, the solution of the steady-state axisymmetric black hole magnetosphere problem is uniquely determined as an eigenfunction problem. As a result, the electric current distribution along open magnetic field lines leads to a total electric current through the black hole jet that does not in general go to zero along the last open magnetic flux surface. As in the case of pulsars, it must therefore close just outside it in the form of a {\em current sheet} (see also Contopoulos, Kazanas \& Fendt 1999). Notice that it is hard to see the current sheet in global GR MHD simulations since these fail to adequately resolve the region immediatly adjacent to the black hole horizon where the inner light cylinder resides. On the other hand, this current sheet is a standard feature of force-free electrodynamic jets (e.g. Tchekhovskoy et al.~2008; Nathanail \& Contopoulos~2014).

It is thus natural to expect that instabilities of various types (Rayleigh-Taylor, kink, sausage, reconnection, anomalous electric current dissipation, etc.) will take place alongside it. Such instabilities are expected to dissipate electromagnetic energy into particle acceleration and radiation, all the way from the equatorial region of the black hole horizon to infinity. We thus conclude this section with the prediction that the EHT will observe energy dissipation along the boundary of the electron-positron black hole jet (see cartoon in Fig.~1). In other words, we expect that the electron-positron black hole jet will be strongly edge brightened.

\section{Magnetic field and wind topology according to the Cosmic Battery}

Current GR MHD numerical simulations of black hole accretion and ejection flows implement an initial magnetic field configuration in which the plasma is threaded by a large scale magnetic field of a definite polarity. In these simulations, the system generates large scale winds and jets, and eventually reaches a so-called MAD state (magnetically arrested disk). Simulations with a random initial magnetic field have also been performed such as the standard and normal evolution-SANE (e.g. Narayan et al.~2015; Chan et al.~2015), but in these the field is introduced only to establish the magnetorotational instability which is needed for accretion to develop. Obviously, SANE simulations do not lead to large scale ejection flows.

Most recently, we performed the first ever GR radiation MHD simulation with an extra electric field component due to the abberation of radiation in the induction equation \cite{Setal17}. This is the GR manifestation of the so-called Cosmic Battery introduced almost 20 years ago by Contopoulos \& Kazanas~(1998). The result of this simulation was that a large scale magnetic field component develops on top of the SANE evolution which eventually has a strong dynamical effect on the accretion flow (it evolves toward a MAD state).

It is important to notice here that, if indeed the magnetic field that threads the black hole is generated by the Cosmic Battery (instead of being brought in from large scales as in all previous GR MHD simulations), it attains a very particular topology:
\begin{enumerate}
\item The magnetic field that is generated around the innermost stable circular orbit (ISCO) is brought to the center by the freely falling accretion flow and threads the black hole horizon. The field returns and closes at larger distances through the viscous and turbulent disk.  
\item The return magnetic field diffuses outward through the disk, all along driving a disk wind (Blandford \& Payne~1982; Contopoulos \& Lovelace~1994; Contopoulos, Kazanas \& Fukumura~2017).
\item At a distance of optical depth unity (this is how far radiation, the driving force of the Cosmic Battery, penetrates in the accretion flow), the magnetic field reverses. We thus expect that a region with weak magnetic field develops at that distance in the disk.
\end{enumerate}

There are several important corrolaries derived from the above magnetic field configuration. Firstly, magnetic field loops will continuously originate around a distance of optical depth unity at the slow rate dictated by the Cosmic Battery (see Contopoulos \& Kazanas~1998 for details). The loops will be stretched in the azimuthal direction by the differential rotation, and as a result they will open up to large axial distances (as in Contopoulos, Nathanail \& Katsanikas~2015). The opening up of the magnetic field may be observable by the EHT at the base of the M87 jet on timescales on the order of days. Secondly, a region with lower magnetic field will develop between the strongly magnetized black hole jet and the surrounding more weakly magnetized disk wind (see cartoon in Fig.~1; see also the last panel of Fig.~3 in Contopoulos et al.~2017 for a clear numerical indication of this effect).

We find hints of the above configuration in the observations of a spine component in the M87 radio jet by Asada et al.~2016. We interpret their observations in the context of the Cosmic Battery as follows. The M87 jet consists of 3 sub-components: an innermost electron-positron Blandford-Znajek black hole jet (spine), an outer (sheath) electron-proton/ion disk wind, and a region in between with weaker radio emission. The electron-positron spine jet becomes unobservable beyond about 1~pc due to radiative cooling and/or Doppler beaming \cite{ANP16}, and what we are left with at large scales is the innermost part of the magnetically driven disk wind. This is what will eventually become the kpc-scale M87 jet.

The latter conclusion (that the kpc-scale jet {\em is not} the electron-positron black hole jet, but is the electron-proton disk wind) is consistent with the observation of Faraday Rotation Measure gradients transverse to the axis of the jet (Christodoulou et al.~2016 and references therein). Such gradients, wherever they are observed, are intrinsic to the jet, and therefore, they indirectly prove that the jet consists of electrons and protons/ions (electron-positron plasmas do not generate Faraday Rotation). We conclude that the kpc-scale jet originates in the disk and not in the black hole. Furthermore, all such detections of Faraday Rotation Measure gradients (firm or tentative) yield axial electric currents that in all cases point away from the central black hole. This particular break of symmetry is consistent with the Cosmic Battery, provided the field polarity associated with the large scale jet is indeed that of the return field that threads the accretion disk around the central black hole (see Christodoulou et al.~2016 for details). 


\section{Summary}

We expect surprises from the EHT. In particular, we expect to observe intermittent variability that is due to particle acceleration electrostatic gaps right above the black hole horizon and not further up in the stagnation surface of GR MHD numerical simulations. We also expect electromagnetic energy dissipation along the boundary of the black hole jet. Such particle acceleration processes have not yet been considered in the numerical simulations used by the EHT Team to interpret the observations. Finally, we expect that a region of weaker magnetic field separates the black hole jet from the magnetically driven disk wind at larger distances, as dictated by the Cosmic Battery. All these elements yield a configuration that consists of an intermittent edge-brightened electron-positron spine jet, a slower and initially weaker electron-proton/ion disk wind at larger cylindrical radii, and a weakly emitting region in between.

We are eagerly awaiting the first EHT observations, especially those of the M87 jet, that will either confirm or disprove our predictions.

\section*{Acknowledgements}
We acknowledge interesting discussions with Drs. George Contopoulos, Christos Efthymiopoulos and Nick Kylafis.



\end{document}